\documentclass{article}

\def\d{\partial}
\def\tr{{\rm tr}}

\begin{document}

\author{Alan D. Rendall\\
Max Planck Institute for Gravitational Physics\\
Am M\"uhlenberg 1\\
14476 Golm\\
Germany
}
\title{Asymptotics of solutions of the Einstein equations with positive
cosmological constant.}

\date{}
\maketitle

\begin{abstract}
A positive cosmological constant simplifies the asymptotics of forever 
expanding cosmological solutions of the Einstein equations. In this
paper a general mathematical analysis on the level of formal power series 
is carried out for vacuum spacetimes of any dimension and perfect fluid
spacetimes with linear equation of state in spacetime dimension four.
For equations of state stiffer than radiation evidence for development
of large gradients, analogous to spikes in Gowdy spacetimes, is found.
It is shown that any vacuum solution satisfying minimal asymptotic 
conditions has a full asymptotic expansion given by the formal series. In 
four spacetime dimensions, and for spatially homogeneous spacetimes of 
any dimension, these minimal conditions can be derived for appropriate 
initial data. Using Fuchsian methods the existence of vacuum spacetimes 
with the given formal asymptotics depending on the maximal number of free 
functions is shown without symmetry assumptions. 
\end{abstract}

\section{Introduction}

Spacetimes with accelerated expansion have come to play an important role in 
cosmology. The accelerating phase may be in the early universe (inflation)
or at the present epoch (quintessence). The simplest way to produce
a model with accelerated expansion which solves the Einstein equations
is to introduce a positive cosmological constant. A good survey article 
on this topic is \cite{straumann}. 

The fact that a positive cosmological constant leads to solutions of the 
Einstein equations with exponential expansion is associated with the term 
'cosmic no hair theorem'. In the following we investigate possibilities
of proving theorems related to these ideas. In the setting of formal 
series a satisfactory answer is obtained for the Einstein equations in 
vacuum or in the presence of a perfect fluid with linear equation of state. 
There are formal series solutions which have the expected asymptotic 
behaviour and which depend on the maximum number of free functions.
This also holds for vacuum spacetimes in higher dimensions. In the case
of even space dimensions it is in general necessary to allow terms with
logarithmic dependence on the expansion parameter. This throws some light
on what is special about three space dimensions. These results are proved
in Section 2.

While most of the results in three space dimensions obtained in Section 2 
confirm the results of \cite{starobinsky}, one new phenomenon was
observed. Evidence is obtained that for fluids with an equation of state
stiffer than that of a radiation fluid inhomogeneous structures can be
formed. This is reminiscent of the formation of spikes near the 
initial singularity in Gowdy spacetimes \cite{rw}.

In Section 3 it is shown that in the vacuum case minimal assumptions on the 
asymptotics in the expanding phase imply that the spacetime has an asymptotic 
expansion of the form already exhibited as formal series. Unfortunately we do 
not know in general how to obtain these minimal assumptions starting from
conditions on initial data. An exception to this is the case of three
space dimensions where it is shown in Section 4 that the minimal assumptions 
can be deduced from results of Friedrich \cite{friedrich86}, 
\cite{friedrich91} based on the conformal method. The minimal assumptions
can also be verified in the case of certain spatially homogeneous vacuum 
spacetimes of any dimension, as shown in Section 5. In particular, there are 
genuine solutions of the Einstein equations whose asymptotics contain
non-vanishing logarithmic terms. In Section 6 Fuchsian methods are applied 
to show the existence of vacuum spacetimes of any dimension with the 
asymptotics given in Section 2 and depending on the maximum number of free 
functions. Finally, Section 7 shows that in a model problem, the wave equation 
on de Sitter space, full information on asymptotics of solutions with 
arbitrary initial data can be obtained.

\section{Perturbative solutions}

A perturbative treatment of four-dimensional vacuum spacetimes with positive
cosmological constant can be found in \cite{starobinsky}. In that paper 
formal solutions are written down without any mathematical
derivation being given. In this section a careful discussion of these
formal power series solutions is presented. The analysis is
generalized to vacuum spacetimes with positive cosmological constant in 
all dimensions. The expansion for perfect fluid spacetimes given in
\cite{starobinsky} is also revisited.

Consider the vacuum Einstein equations with cosmological constant 
$\Lambda$ for a spacetime of dimension $n+1$ with $n\ge 2$. An $n+1$ 
decomposition with lapse equal to one and vanishing shift results
in the constraint equations
\begin{eqnarray}
R-k^{ab}k_{ab}+(\tr k)^2=2\Lambda\label{hamconstraint}           \\
\nabla_a k^a{}_b-\nabla_b (\tr k)=0\label{momconstraint}
\end{eqnarray}
and the evolution equation
\begin{equation}\label{evolution}
\d_t k^a{}_b=R^a{}_b+(\tr k)k^a{}_b-\frac{2\Lambda}{n-1}\delta^a{}_b
\end{equation}
Here $g_{ab}$ is the spatial metric with Ricci tensor $R_{ab}$ and scalar 
curvature $R$, $k_{ab}$ is the second fundamental form and indices are raised 
and lowered using $g_{ab}$ and its inverse. Let $\sigma^a{}_b$ be the 
trace-free part of the second fundamental form and $\tilde R^a{}_b$ the 
trace-free part of the spatial Ricci tensor and define the following 
quantities:
\begin{eqnarray}
\tilde E^a{}_b&=&\d_t\sigma^a{}_b-[\tilde R^a{}_b+(\tr k)\sigma^a{}_b]  \\
E&=&\d_t (\tr k)-\left[R+(\tr k)^2-\frac{2n\Lambda}{n-1}\right]      \\
C&=&R-k^{ab}k_{ab}+(\tr k)^2-2\Lambda                     \\
C_a&=&\nabla_b k^b{}_a-\nabla_a (\tr k)
\end{eqnarray}
The Einstein equations are equivalent to the vanishing of the 
evolution quantities $\tilde E^a{}_b$ and $E$ and the constraint quantities
$C$ and $C_a$. These quantities are linked by the following consistency
conditions
\begin{eqnarray}
\d_t C&=&2(\tr k)C-2\nabla^a C_a-2\sigma^a{}_b\tilde E^b{}_a
+2(1-1/n)(\tr k)E\label{hamconsistency} \\
\d_t C_a&=&(\tr k)C_a-(1/2)\nabla_a C+\nabla_b\tilde E^b{}_a-(1-1/n)\nabla_a E
\label{momconsistency}
\end{eqnarray}

To do a perturbative analysis of equations 
(\ref{hamconstraint})-(\ref{evolution}) consider the formal power series
\begin{equation}\label{formal} 
g_{ab}=e^{2Ht}(g^0_{ab}+g^1_{ab}e^{-Ht}+g^2_{ab}e^{-2Ht}
+g^3_{ab}e^{-3Ht}+\ldots)
\end{equation}
where $H$ is a constant. The $n+1$ form of the Einstein equations are imposed 
on this expression in a suitable sense. It turns out that for consistency it 
is necessary to choose $H=\sqrt{2\Lambda/(n(n-1))}$ and so, in particular, 
$\Lambda$ must be positive. Products of formal series are defined in the
obvious way that the terms in the individual series are multiplied and the
resulting terms with the same power of $e^{-Ht}$ collected. The derivatives 
of a formal series with respect to the space and time variables are  defined 
via term by term differentiation. In order to impose the Einstein equations 
it is also necessary to have a definition of the inverse $g^{ab}$ of the 
formal power series metric $g_{ab}$. This can be done uniquely by requiring 
that the relation $g_{ab}g^{bc}=\delta_a^c$ holds. This allows the 
coefficient of order $m$ in the series for $g^{ab}$ to be expressed in
terms of the coefficients in the series for $g_{ab}$ up to order $m$. Setting
$g^0_{ab}=\delta_{ab}$ and $g^m_{ab}=0$ for $m>0$ gives an exact solution of
the Einstein equations. In the case $n=3$ it is the de Sitter solution 
(\cite{he}, p. 125).

Given any tensor $T$, let $(T)_m$ denote the coefficient of $e^{-mHt}$ in 
the expansion of $T$. With this notation $(g_{ab})_m=g^{m+2}_{ab}$. It
follows from (\ref{formal}) that $(R^a{}_b)_m=0$ for $m=0$ and $m=1$.   
It also follows directly from (\ref{formal}) that $(\tr k)_0=-nH$
and $(\sigma^a{}_b)_0=0$. This is consistent with the vanishing of the
coefficients of all evolution and constraint quantities for $m=0$.
The vanishing of $(E)_1$ and $(\tilde E^a{}_b)_1$ implies that 
$(\tr k)_1=0$ and $(\sigma^a{}_b)_1=0$. It follows that $(C)_1=0$
and $(C_a)_1=0$ and this ensures the consistency of the series up to 
order $m=1$. Using the relation $\d_t g_{ab}=-2g_{ac}k^c{}_b$ shows
that $g^1_{ab}=0$ and this in turn implies that $(R^a{}_b)_3=0$.

The relations between coefficients for $m\ge 2$ will now be written down.
The summation indices $p$ and $q$ in the following formulae are assumed
to be no less than two. The evolution equations (\ref{evolution}) imply the 
recursion relations
\begin{equation}\label{shearm}
(n-m)(\sigma^a{}_b)_m=H^{-1}\left[\sum_{p+q=m}(\sigma^a{}_b)_p(\tr k)_q+
(\tilde R^a{}_b)_m\right]
\end{equation}
and
\begin{equation}\label{meanm}
(2n-m)(\tr k)_m=H^{-1}\left[\sum_{p+q=m}(\tr k)_p(\tr k)_q+(R)_m\right]
\end{equation}
The Hamiltonian constraint (\ref{hamconstraint}) gives 
\begin{equation}\label{hamm}
(2n-2)(\tr k)_m=H^{-1}\left\{\sum_{p+q=m}[-(k^a{}_b)_p(k^b{}_a)_q
+(\tr k)_p(\tr k)_q]
+(R)_m\right\}
\end{equation}
and the momentum constraint (\ref{momconstraint}) gives
\begin{equation}\label{momm}
\nabla^0_a (k^a{}_b)_m-\nabla^0_b(\tr k)_m=\sum_{p+q=m}[(\Gamma^c_{ab})_p
(k^a{}_c)_q-(\Gamma^a_{ac})_p (k^c{}_b)_q]
\end{equation}
Here $\nabla^0$ is the covariant derivative associated to $g^0_{ab}$.
The consistency conditions (\ref{hamconsistency}) and (\ref{momconsistency})
relating evolution and constraint quantities imply
that if $(E)_k$, $(\tilde E^a{}_b)_k$, $(C)_k$ and $(C_a)_k$ all vanish for
$k\le m-1$ then
\begin{eqnarray}
(2n-m)(C)_m&=&-2(n-1)(E)_m\label{conham}  \\
(n-m)H(C_a)_m&=&-(1/2)\nabla_a(C)_m+\nabla^0_b(\tilde E^b{}_a)_m
-(1-1/n)\nabla_a(E)_m\label{conmom}
\end{eqnarray}

Consider first the case $n=3$. The form (\ref{formal}) of the series for 
$g_{ab}$, taking account of the vanishing of the coefficient $g^1_{ab}$, is 
contained in \cite{starobinsky}. The following theorem formalizes some of 
the statements in \cite{starobinsky}. Here smooth means $C^\infty$.

\vskip 10pt\noindent
{\bf Theorem 1} Let $A_{ab}$ be a smooth three-dimensional Riemannian metric 
and $B_{ab}$ a smooth symmetric tensor which satisfies $A^{ab}B_{ab}=0$
and $\nabla^a B_{ab}=0$, where the covariant derivative is that 
associated to $A_{ab}$. Then there exists a unique formal power series
solution of the vacuum Einstein equations with cosmological constant
$\Lambda>0$ of the form (\ref{formal}) with $g^0_{ab}=A_{ab}$ and 
$g^3_{ab}=B_{ab}$. The coefficients $g^m_{ab}$ are smooth. 

\noindent
{\bf Proof} The coefficients $(k^a{}_b)_m$ determine the coefficients 
$(g_{ab})_m$ recursively. For substituting (\ref{formal}) into the relation
$\d_t g_{ab}=-2g_{ac}k^c{}_b$ gives 
\begin{equation}
mHg^m_{ab}=2g^0_{ac}(k^c{}_b)_m+2\sum_{p+q=m}g^p_{ac}(k^c{}_b)_q
\end{equation}
and hence an equation which expresses 
$(g_{ab})_{m-2}$ in terms of $(k^a{}_b)_m$ and lower order terms
for any $m\ge 2$. Thus in order to prove the theorem it is enough to
show that equations (\ref{shearm})-(\ref{momm}) determine the 
coefficients $(k^a{}_b)_m$ uniquely and that when the coefficients have 
been fixed in this way all the equations (\ref{shearm})-(\ref{momm})
are satisfied. The coefficient $(k^a{}_b)_m$ is determined by (\ref{shearm})
and (\ref{meanm}) for all $m\ge 2$ except $m=3$ and $m=6$. The coefficient
$(k^a{}_b)_3$ is determined by using the condition that $g^3_{ab}=B_{ab}$.
The coefficient $(\sigma^a{}_b)_6$ is determined by (\ref{shearm}) while 
$(\tr k)_6$ is determined by (\ref{hamm}). By 
construction the evolution equation (\ref{meanm}) is satisfied for 
all values of $m$ except possibly $m=3$ and $m=6$ while (\ref{shearm})
is satisfied except possibly for $m=3$. The fact that $B_{ab}$ has zero trace
ensures that (\ref{meanm}) is satisfied while (\ref{shearm}) is 
automatic for $m=3$. It will now be shown by induction that 
(\ref{shearm})-(\ref{momm}) hold for all $m$. Except for $m=6$ only
(\ref{hamm}) and (\ref{momm}) need to be verified. The equations 
(\ref{shearm})-(\ref{momm}) hold for $m=1$. For $2\le m\le 5$ the inductive
step from $m-1$ to $m$ can be carried out as follows. When $m\ne 3$ the
consistency condition (\ref{conham}) shows that $(C)_m=0$ and then the
consistency condition (\ref{conmom}) shows that $(C_a)_m=0$. That $(C)_3$
and $(C_a)_3$ are zero follows from the conditions on $B_{ab}$ in the
hypotheses of the theorem. Knowing that the equations 
(\ref{shearm})-(\ref{momm}) hold for $m\le 5$ implies using (\ref{conham})
that $(E)_6=0$. By construction $(C)_6=0$ and then it is straightforward
to obtain $(C_a)_6=0$ from (\ref{conmom}). For $m\ge 7$ we can proceed as 
for $2\le m\le 5$.
This completes the proof.

\vskip 10pt\noindent
{\bf Remark} If $P_{ab}$ and $P$ denote the Ricci tensor and Ricci scalar 
of $g^0_{ab}$ respectively, then $g^{2}_{ab}=H^{-2}(P_{ab}-(1/4) Pg^0_{ab})$,
a relation given in \cite{starobinsky}. 

The theorem just proved can be generalized directly to all larger odd values
of $n$, as will now be shown.

\vskip 10pt\noindent
{\bf Theorem 2} Let $A_{ab}$ be a smooth $n$-dimensional Riemannian metric 
with $n$ odd and $B_{ab}$ a smooth symmetric tensor which satisfies 
$A^{ab}B_{ab}=0$ and $\nabla^a B_{ab}=0$, where the covariant derivative is 
that associated to $A_{ab}$. Then there exists a unique formal power series
solution of the vacuum Einstein equations with cosmological constant
$\Lambda>0$ of the form (\ref{formal}) with $g^0_{ab}=A_{ab}$ and 
$g^n_{ab}=B_{ab}$. The coefficients $g^m_{ab}$ are smooth.

\noindent
{\bf Proof} Let $s$ be an integer such that $2s+1<n$ and $(k^a{}_b)_m=0$
for all odd $m$ with $m\le 2s-1$. If follows that $g^m_{ab}=0$ for all
odd $m$ with $m\le 2s-1$ and that $(g^{ab})_m=0$ for all odd $m$ with
$m\le 2s+1$. Putting this information into the Ricci tensor shows that 
$(R^a{}_b)_m=0$ for all odd $m$ with $m\le 2s+1$. Then (\ref{shearm})
and (\ref{meanm}) imply that $(k^a{}_b)_{2s+1}=0$. It can then be proved
by induction that $(k^a{}_b)_m=0$ vanishes for all odd $m$ with $m<n$.
{}From this point on the proof is very similar to that of the previous
theorem. The coefficients $(k^a{}_b)_m$ are uniquely determined for all 
values of $m$ except $m=n$ and $m=2n$. The coefficient
$(k^a{}_b)_n$ is determined by using the condition that $g^n_{ab}=B_{ab}$.
The coefficient $(\sigma^a{}_b)_{2n}$ is determined by (\ref{shearm}) while 
$(\tr k)_{2n}$ is determined by (\ref{hamm}). By 
construction the evolution equation (\ref{meanm}) is satisfied for 
all values of $m$ except possibly $m=n$ and $m=2n$ while (\ref{shearm})
is satisfied except possibly for $m=n$. The fact $B_{ab}$ has zero trace
ensures that (\ref{meanm}) is satisfied while (\ref{shearm}) is 
automatic for $m=n$. These statements make use of the fact that the odd
order coefficients of $k^a{}_b$ of order less than $n$ vanish.
It will now be shown by induction that 
(\ref{shearm})-(\ref{momm}) hold for all $m$. Except for $m=2n$ only
(\ref{hamm}) and (\ref{momm}) need to be verified. The equations 
(\ref{shearm})-(\ref{momm}) hold for $m=1$. For $2\le m\le 2n-1$ the inductive
step from $m-1$ to $m$ can be carried out as follows. When $m\ne n$ the
consistency condition (\ref{conham}) shows that $(C)_m=0$ and then the
consistency condition (\ref{conmom}) shows that $(C_a)_m=0$. That $(C)_n$
and $(C_a)_n$ are zero follows from the conditions on $B_{ab}$ in the
hypotheses of the theorem. Knowing that the equations 
(\ref{shearm})-(\ref{momm}) hold for $m\le 2n-1$ implies using (\ref{conham})
that $(E)_{2n}=0$. By construction $(C)_{2n}=0$ and then it is straightforward
to obtain $(C_a)_{2n}=0$. For $m\ge 2n+1$ we can proceed as for 
$2\le m\le 2n-1$. This completes the proof.

\vskip 10pt
The case where $n$ is even is more complicated. The form (\ref{formal}) of
the metric must be generalized to
\begin{equation}\label{logformal}
g_{ab}=e^{2Ht}(g^0_{ab}+\sum_{m=0}^\infty\sum_{l=0}^{L_m}(g_{ab})_{m,l}
t^le^{-mHt})
\end{equation}
where $L_m$ is a non-negative integer for each $m$ and $L_m=0$ for $m<n$.
Given any tensor $T$ with an expansion of the above type let $(T)_{m,l}$ 
denote the coefficient of $t^le^{-mHt}$. As before when manipulating series
they are differentiated term by term. The recursion relations for the
expansion coefficients coming from the evolution equations generalize as 
follows, where the terms not written out explicitly are lower order in the 
sense that they can be expressed in terms of the coefficients of $k^a{}_b$ 
with $m$ smaller:
\begin{eqnarray}
(n-m)(\sigma^a{}_b)_{m,l}+H^{-1}(l+1)(\sigma^a{}_b)_{m,l+1}=\ldots
\label{logshearm}  \\
(2n-m)(\tr k)_{m,l}+H^{-1}(l+1)(\tr k)_{m,l+1}=\ldots\label{logmeanm} 
\end{eqnarray}
The terms on the right hand side not written out are obtained from
the terms on the right hand side of equations (\ref{shearm}) and (\ref{meanm})
if the indices $m$, $p$ and $q$  are replaced by the pairs $(m,l)$, $(p,l_1)$
and $(q,l_2)$, summing over $l_1+l_2=l$. The recursion relations implied by 
the constraints are identical except for the addition of an extra index $l$. 
In a similar way, the consistency conditions lead to
\begin{eqnarray}
(2n-m)(C)_{m,l}+H^{-1}(l+1)(C)_{m,l+1}&=&-2(n-1)(E)_{m,l}\label{logconham}  \\
(n-m)H(C_a)_{m,l}+H^{-1}(l+1)(C_a)_{m,l+1}&=&
-(1/2)\nabla_a(C)_{m,l}\nonumber   \\
+\nabla^0_b(\tilde E^b{}_a)_{m,l}
&-&(1-1/n)\nabla_a(E)_{m,l}\label{logconmom}
\end{eqnarray}
assuming that $(E)_{k,l}$. $(\tilde E^a{}_b)_{k,l}$, $(C)_{k,l}$ and 
$(C_a)_{k,l}$ vanish whenever $k\le m-1$.  

By using the above relations it is possible to express 
$g^{ab}_0(g_{ab})_{n-2,0}$ as a function of $g^0_{ab}$ and its spatial 
derivatives. We denote this schematically by 
$g^{ab}_0(g_{ab})_{n-2,0}=Z(g^0)$. Similarly, it is possible
to write ${}^0\nabla^a (g_{ab})_{n-2,0}=\tilde Z_b(g^0)$. 

\vskip 10pt\noindent
{\bf Theorem 3} Let $A_{ab}$ be a smooth $n$-dimensional Riemannian metric 
and $B_{ab}$ a smooth symmetric tensor which satisfies 
$A^{ab}B_{ab}=Z(A)$ and $\nabla^a B_{ab}=\tilde Z_b(A)$, where the covariant 
derivative is that associated to $A_{ab}$. Then there exists a unique formal 
series solution of the vacuum Einstein equations with cosmological 
constant $\Lambda>0$ of the form (\ref{logformal}) with $g^0_{ab}=A_{ab}$ 
and $(g_{ab})_{n-2,0}=B_{ab}$. The coefficients $(g_{ab})_{m,l}$ are smooth.

\noindent
{\bf Proof} As a general principle, when determining coefficients for
fixed $m$ we start from $l=L_m$ and proceed to successively lower values
of $l$. If $n$ is odd then the existence follows from Theorem 2. Uniqueness
for $n$ odd in the wider class being considered in this theorem in 
comparison with Theorem 2 is obtained by a straightforward extension of the 
argument given in the proof of the latter. Consider
now the case where $n$ is even. By analogy with the proof of Theorem 2
it can be shown that $(k^a{}_b)_{m,l}=0$ for $m$ odd. For $m<n$ the 
coefficients are uniquely determined. The assumption that $L_m=0$ in
this range is thus unavoidable.The coefficients $(\tr k)_{n,l}$ are
uniquely determined by (\ref{logmeanm}) and vanish for $l>0$. The
coefficients $(\sigma^a{}_b)_{n,l}$ are uniquely determined for $l\ge 1$
and vanish for $l>1$. The choice of $B_{ab}$ determines 
$(\sigma^a{}_b)_{n,0}$. For $n<m<2n$ equations (\ref{logshearm}) and
(\ref{logmeanm}) determine $(k^a{}_b)_{m,l}$. Equation (\ref{logshearm}) is 
used to determine $(\sigma^a{}_b)_{2n,l}$ while the analogue of (\ref{hamm}) 
is used to determine $(\tr k)_{2n,l}$. For $m>2n$ (\ref{logshearm}) and 
(\ref{logmeanm}) can be used again. That all field equations are satisfied
at all orders can be proved much as in the proof of Theorem 2, always 
proceeding in the direction of decreasing $l$ for each fixed $m$.

\vskip 10pt
A question left open by Theorem 3 is whether it can ever happen that
any of the coefficients with $l>0$ are non-zero. This is equivalent to
the question whether $(\sigma^a{}_b)_{n,1}$ is ever non-zero. It
follows from the proof of the theorem that this coefficient is 
uniquely determined by $g^0_{ab}$. In the case $n=2$ the coefficient 
of interest vanishes due to the fact that the Ricci tensor of a
two-dimensional metric is automatically traceless. For all even 
dimensions greater than two there are choices of $g^0_{ab}$ for which
$(\sigma^a{}_b)_{n,1}$ does not vanish. In fact this is the generic case.
The coefficient of interest can be written as a polynomial expression
in $H^{-1}$. If there were no logarithmic terms for a given choice of 
$g^0_{ab}$ then all terms in this polynomial would have to vanish.
The coefficient of $H^{-n+1}$, which is the most negative power of $H$ 
occurring, is a non-zero constant times $\tilde P^a{}_b (\tr P)^{k-1}$,
where $k=n/2$ and $\tilde P^a{}_b$ and $\tr P$ are the tracefree part and
trace of the Ricci tensor of $g^0_{ab}$. There are only two ways in 
which this coefficient can vanish. Either the scalar curvature of
$g^0_{ab}$ vanishes identically or $g^0_{ab}$ is an Einstein metric.
A necessary condition for the absence of logarithmic terms has now been
given but it is unlikely to be sufficient. The coefficients of other powers 
of $H$ have to be taken into account in order to decide this issue. 

In \cite{starobinsky} the expansions obtained for vacuum spacetimes were
extended to the case of a perfect fluid with pressure proportional to
energy density. Formalizing these considerations leads to a theorem 
generalizing Theorem 1 above. The notation here is as follows: $\rho=T^{00}$,
$j^a=T^{0a}$ and $S^{ab}=T^{ab}$. The proper energy density and pressure of 
the fluid are denoted by $\mu$ and $p$ respectively, so that
\begin{equation}
T^{\alpha\beta}=(\mu+p)u^\alpha u^\beta+pg^{\alpha\beta}
\end{equation}
The equation of state is taken to be $p=(\gamma-1)\mu$ with $1\le\gamma<2$.
In the case with matter evolution 
and constraint quantities can be defined by
\begin{eqnarray}
\tilde E^a{}_b&=&\d_t\sigma^a{}_b-[\tilde R^a{}_b+(\tr k)\sigma^a{}_b
-8\pi\tilde S^a{}_b]                                                    \\
E&=&\d_t (\tr k)-[R+(\tr k)^2+4\pi\tr S-12\pi\rho-3\Lambda]             \\
C&=&R-k^{ab}k_{ab}+(\tr k)^2-16\pi\rho-2\Lambda                     \\
C_a&=&\nabla_b k^b{}_a-\nabla_a (\tr k)-8\pi j_a
\end{eqnarray}
so that their vanishing is equivalent to the Einstein equations. These 
satisfy the consistency conditions (\ref{hamconsistency}) and
(\ref{momconsistency}) as in the vacuum case. The components of the 
energy-momentum tensor can be expressed in terms of the fundamental
fluid variables as follows:
\begin{eqnarray}
\rho&=&\mu(1+\gamma|u|^2)                        \\
j^a&=&\gamma\mu(1+|u|^2)^{1/2}u^a          \\
S^a{}_b&=&\mu[\gamma u^au_b+(\gamma-1)\delta^a_b]
\end{eqnarray} 
where $|u|^2=g_{ab}u^au^b$. The following relations will be useful:
\begin{eqnarray}
\d_t\rho-(\tr k)\rho-\frac13(\tr k)\tr S&=&-\nabla_a j^a+\sigma^a{}_b S^b{}_a
\\
\d_t j^a-\frac53 (\tr k) j^a&=&-\nabla^b S^a{}_b+2\sigma^a{}_b j^b
\end{eqnarray}

It is possible to express $\mu$ and $u^a$ in terms of $\rho$ and $j^a$. To
see this note first that $|j|=\gamma\mu(1+|u|^2)^{1/2}|u|$ and that
as a consequence:
\begin{equation}
|j|^2/\rho^2=\gamma^2(1+|u|^2)|u|^2/(1+\gamma |u|^2)^2
\end{equation}
If $f(x)=\gamma^2 x^2(1+x^2)(1+\gamma x^2)^{-2}$ then
$f'(x)=2\gamma^2 x(1+\gamma x^2)^{-3}(1+(2-\gamma)x^2)>0$. It follows that
the mapping from the interval $[0,\infty)$ to the interval $[0,1)$
defined by $f$ is invertible and $|u|^2$ can be expressed as a smooth 
function of $|j|^2/\rho^2$ for $\rho>0$. Since $\mu$ can be expressed as a 
smooth function of $\rho$ and $|u|^2$ it follows that it is a smooth function 
of $\rho$ and $j^a$. Similarly the fact that $u^a$ can be expressed as a 
smooth function of $\mu$, $|u|^2$ and $j^a$ implies that $u^a$ is a smooth 
function of $\rho$ and $j^a$.

Next Theorem 1 will be generalized to the case with perfect fluid. The
solution is sought as a formal series where each tensor occurring is 
written as a sum of exponentials. The exponents are taken from an
increasing sequence of real numbers $M=\{m_i\}$ which tends 
to infinity as $i\to\infty$. The solution is of the form
\begin{eqnarray}
g_{ab}&=&\sum_{m_i\in M}(g_{ab})_{m_i}e^{-m_iHt}\label{fluidformal}     \\
\mu&=&\sum_{m_i\in M}(\mu)_{m_i}e^{-m_iHt} \nonumber                    \\
u^a&=&\sum_{m_i\in M}(u^a)_{m_i}e^{-m_iHt} \nonumber
\end{eqnarray}
Let integers $k_1$, $k_2$, $k_3$ and $k_4$ be defined as follows. For 
$\gamma\le 4/3$ we have $k_1=3\gamma$, $k_2=5-3\gamma$, $k_3=3\gamma$,
$k_4=5$ while for $\gamma\ge 4/3$ we have $k_1=2\gamma/(2-\gamma)$, 
$k_2=(6-4\gamma)/(2-\gamma)$, $k_3=4$ and $k_4=5$. In order to
organize the coefficients it is useful to define the relative order
$\tilde m_i$ of a coefficient of order $m_i$. For the quantities
$g_{ab}$, $k^a{}_b$, $\mu$, $u^a$, $\rho$ and $j^a$ these are defined by 
$\tilde m_i=m_i+2$, $\tilde m_i=m_i$, $\tilde m_i=m_i-k_1$, 
$\tilde m_i=m_i-k_2$, $\tilde m_i=m_i-k_3$ and $\tilde m_i=m_i-k_4$
respectively. 

\vskip 10pt\noindent
{\bf Theorem 4} Let $A_{ab}$ be a smooth three-dimensional Riemannian metric 
and $B_{ab}$ a smooth symmetric tensor, $\mu_0$ a smooth positive real-valued 
function and $u^a_0$ a smooth vector field. Suppose that 
$A^{ab}B_{ab}=-(8\pi/3H^2)\mu_0$ for $\gamma=1$, $A^{ab}B_{ab}=0$ for 
$\gamma>1$ and $\nabla^a B_{ab}=\nabla_b(A^{ac}B_{ac})+(16\pi\gamma/3H)\mu_0 
A_{bc}u^c_0$ 
where the covariant derivative is that associated to $A_{ab}$. If $\gamma>4/3$
suppose furthermore that $u^a$ is nowhere vanishing. Then there exists a 
unique 
formal power series solution of the Einstein-Euler equations with cosmological 
constant $\Lambda>0$ and equation of state $p=(\gamma-1)\rho$, $1\le\gamma <2$,
of the form (\ref{fluidformal}) with $(g_{ab})_{-2}=A_{ab}$, 
$(g_{ab})_1=B_{ab}$, $(\mu)_{k_1}=\mu_0$ and $(u^a)_{k_2}=u^a_0$. The 
coefficients of the series are smooth. They satisfy $(\mu)_{m_i}=0$ for 
$m_i<k_1$, $(u^a)_{m_i}=0$ for $m_i<k_2$ and, except for $m_i=-2$,  the 
coefficient $(g_{ab})_{m_i}$ vanishes for $m_i<0$.

\noindent
{\bf Proof} Consider a formal series solution whose coefficients vanish in the
ranges indicated in the statement of the theorem. With the given values for 
$k_1$ and $k_2$ the matter terms do not contribute to the equations for the 
coefficients of $k^a{}_b$ below order three and thus all statements made about 
these coefficients in the vacuum case can be taken over without change.
This follows from the fact that $\rho$, $S^a{}_b$ and $j_a$ are all
$O(e^{-3Ht})$. The exponent in this estimate can be improved except in the 
case of $\rho$ with $\gamma=1$ and in the case of $j_a$ with general $\gamma$.

The proof splits into several cases. Suppose first that 
$1\le\gamma<1/3$. Then $|u|=o(1)$ and so in leading order $\rho=\mu$,
$j^a=\gamma\mu |u|u^a$ and $S^a{}_b=(\gamma-1)\mu\delta^a_b$. Thus the 
following relations are obtained:
\begin{eqnarray}
(m_i-3\gamma)(\rho)_{m_i}&=&\ldots  \\
(m_i-5)(j^a)_{m_i}&=&\ldots
\end{eqnarray}
The terms not written out explicitly are lower order in the sense that
they are combinations of terms of lower relative order than $\tilde m_i$.
There is one subtlety involved in showing this. In the case $\gamma=1$ the 
expression $\nabla^b S^a{}_b$ gives rise to a term which, looking at the 
exponents, is not lower order. However the coefficient of this term contains a
factor $\gamma-1$ and so the term vanishes for $\gamma=1$. The Einstein 
equations give:
\begin{equation}
(3-m_i)(\sigma^a{}_b)_{m_i}=\ldots
\end{equation}
and
\begin{equation}
(6-m_i)(\tr k)_{m_i}=-12\pi H^{-1}(\rho)_{m_i}+\ldots
\end{equation}
The terms on the right hand side of the last two equations not written out 
explicitly are lower order. The one explicit term on the right hand side of 
the last equation is also lower order except in the case $\gamma=1$. The
energy-momentum quantities $\rho$ and $j^a$ are linked to the matter
quantities $\mu$ and $u^a$ by the relations
\begin{eqnarray}
(\mu)_{m_i}&=&{\rho}_{m_i}+\ldots               \\
(u^a)_{m_i}&=&\gamma^{-1}(\rho^{-1}j^a)_{m_i}+\ldots
\end{eqnarray}
Fix a value of $\tilde m_i$ and suppose that all coefficients with lower
relative order have been determined. Consider the equations for $(\rho)_{m_i}$
and $(u^a)_{m_i}$. These coefficients are determined uniquely unless
$\tilde m_i=0$ and if $\tilde m_i<0$ they vanish. When $\tilde m_i=0$ 
they are determined by the conditions on $(\mu)_{k_1}$ and $(u^a)_{k_2}$
in the hypotheses of the theorem and the equations relating $\rho$ and
$j^a$ to $\mu$ and $u^a$. The latter relations also fix the coefficients
of $\mu$ and $u^a$ of the given relative order when $\tilde m_i>0$. Next
consider the equations for $(\tr k)_{m_i}$ and $(k^a{}_b)_{m_i}$. By what 
has been said above we may assume that $m_i\ge 3$. The unique determination
of the coefficients of the given relative order can be shown using the same
procedure as in the vacuum case. The additional terms are either already
of lower relative order, and hence known, or have been determined in the
preceeding discussion of the matter equations. By induction on $i$ it can
be concluded that all coefficients are uniquely determined. The fact that
all field equations are satisfied can be shown much as in the vacuum
case since the compatibility conditions are identical.

Now consider the case $4/3<\gamma<2$. The assumption that $u_0^a$ is
nowhere vanishing implies in this case that $|u|^{-1}=o(1)$ and in 
leading order $\rho=\gamma\mu |u|^2$, $j^a=\gamma\mu |u|u^a$ and
$S^a{}_b=\gamma\mu u^a u_b$. The following relations are obtained:
\begin{eqnarray}
(m_i-4)(\rho)_{m_i}&=&\ldots  \\
(m_i-5)(j^a)_{m_i}&=&\ldots
\end{eqnarray}
The Einstein equations give the same relations as in the previous case.
For $4/3<\gamma<2$ the energy-momentum quantities $\rho$ and $j^a$ are 
linked to the matter quantities $\mu$ and $u^a$ by the relations
\begin{eqnarray}
(\mu)_{m_i}&=&(2-\gamma)^{-1}(\rho(1-|j|^2/\rho^2))_{m_i}+\ldots         \\
(u^a)_{m_i}&=&((2-\gamma)/\gamma)^{1/2}(\rho^{-1}(1-|j|^2/
\rho^2)^{-1/2}j^a)_{m_i}+\ldots
\end{eqnarray}
Using these facts we can proceed as in the case $1\le\gamma<4/3$.

Consider finally the case $\gamma=4/3$ where $|u|$ tends to a finite 
limit, in general non-zero, as $t\to\infty$. The difference in comparison
to the cases already treated is that the relations between $\rho$ and $j^a$
on the one hand and $\mu$ and $u^a$ on the other hand cannot be
inverted explicitly in leading order. However the fact, shown above, that 
the relevant mappings are invertible has an equivalent on the level of
formal power series. For if $f$ is a smooth function between open subsets of 
Euclidean spaces then $f(x+y)$ can be written formally in terms of a Taylor 
series about $x$. The resulting expression contains the derivatives of $f$ 
evaluated at $x$ multiplied by powers of $y$. If $y$ is replaced by a formal 
power series without constant term then a well-defined formal power series 
for $f(x+y)$ is obtained. Thus the same method can be applied as in the 
previous cases, allowing the proof of the theorem to be completed.

\vskip 10pt
A case which has been excluded in the above theorem is that where 
$\gamma>4/3$ and $u^a$ may vanish somewhere. In that case the kind of
series which has been assumed in the theorem is not consistent. For
if it is assumed that an expansion of this kind is possible this leads
to different rates of decay for certain quantities, for instance $\mu$,
depending on whether $|u|$ does or does not vanish. As a consequence
$\nabla_a\mu/\mu$ will be unbounded as $t$ tends to infinity although
$\mu$ is nowhere zero. This contradicts the assumptions which have been 
made. The situation is reminiscent of the spikes observed near the 
initial singularity in Gowdy spacetimes \cite{rw} and so we may
speculate that in reality inhomogeneous features develop in $\mu$
so that the density contrast blows up as $t\to\infty$. This behaviour
for $\gamma>4/3$ is not consistent with the usual picture in inflationary 
models where the density contrast remains bounded at late times. The
issue deserves to be investigated further.

It is interesting to ask whether the expansions for a fluid presented 
here can be extended to the case of collisionless matter. If they can 
then the result probably resembles that for dust. Limited expansions
in some special cases are already known \cite{lee}, \cite{tchapnda}.

Note that the analysis of vacuum spacetimes in this section has a close
analogue for Riemannian (i.e. positive definite) metrics. A solution of
the Einstein equations with positive cosmological constant in the 
Lorentzian case corresponds to an Einstein metric with negative Einstein
constant in the Riemannian case. The equations obtained for a positive
definite metric are
\begin{eqnarray}
-R-k^{ab}k_{ab}+(\tr k)^2&=&-(n-1)K                  \\
\nabla_a k^a{}_b-\nabla_b (\tr k)&=&0                \\ 
\d_t k^a{}_b&=&-R^a{}_b+(\tr k)k^a{}_b+2K\delta^a{}_b
\end{eqnarray}
where $K$ is the Einstein constant, i.e. the $n+1$-dimensional metric 
satisfies $R_{\alpha\beta}=Kg_{\alpha\beta}$. Asymptotic expansions for this
case have been investigated in the literature on Riemannian geometry 
\cite{fefferman} and string theory \cite{skenderis}. 

\section{From minimal to full asymptotics} 

In the last section consistent formal asymptotic expansions were exhibited
for a number of problems. In this section it is shown that minimal 
information about the asymptotics implies the full expansions given in
the last section. For simplicity we restrict consideration to the vacuum
case. The following lemma will be used:

\vskip 10pt\noindent
{\bf Lemma 1} Consider an equation of the form
\begin{equation}\label{decay}
\d_t u+k u=\sum_{m,l} v_{m,l} t^l e^{-mt}+O(e^{-jt})
\end{equation}
for a vector-valued function $u(t)$, where $j\ne k$ and $m<j$ in the sum. 
Then there are coefficients $u_{m,l}$, $m<j$, such that
\begin{equation}\label{series}
u=\sum_{m,l} u_{m,l} t^l e^{-mt}+O(e^{-jt})
\end{equation}
If (\ref{decay}) may be differentiated term by term with respect to $t$
as often as desired the same is true of (\ref{series}).

\noindent
{\bf Proof} Note first that $\d_t (e^{kt}u)$ is equal to a sum of explicit
terms with a term of order $e^{(k-j)t}$. Each of the explicit 
terms has an explicit primitive which is a sum of terms of the same 
general form and the same value of $m$ but in general several values of
$l$. Thus we can absorb these terms into the time derivative and write
\begin{equation}
\d_t(e^{kt}(u-\sum_{m,l} u_{m,l} t^l e^{-mt}))=O(e^{(k-j)t})
\end{equation}
with $m<j$ in the sum. If $j<k$ we can integrate this relation
directly to get the desired result. If $j>k$ then the expression which
is differentiated with respect to time converges to a limit as 
$t\to\infty$, which can be called $u_{k,0}$. This gives the desired result
in the latter case.  

If the assumption on time 
derivatives is satisfied then $\d_t u$ satisfies an equation of the same 
form as that
satisfied by $u$. Hence $\d_t u$ has an asymptotic expansion
\begin{equation}
\d_t u=\sum_{m,l} w_{m,l} t^l e^{-mt}+O(e^{-jt})
\end{equation}
Integrating this from $t_0$ to $t$ and using (\ref{series}) gives
\begin{equation}
\sum_{m,l}\int_{t_0}^t w_{m,l}s^le^{-ms} ds =C+\sum_{m,l}u_{m,l}t^le^{-mt}
+O(e^{-jt})
\end{equation}
for a constant $C$. It follows that the coefficients $w_{m,l}$ are obtained 
from $u_{m,l}$ by term by term differentiation. This process can be repeated 
for higher order derivatives with respect to $t$.

\vskip 10pt\noindent
{\bf Remark} If the quantities in (\ref{decay}) depend smoothly on a
parameter and the equation may be differentiated term by term with
respect to the parameter then the same is true for the solution.

\vskip 10pt\noindent
{\bf Theorem 5} Let a solution of the vacuum Einstein equations with 
cosmological constant $\Lambda>0$ in $n+1$ dimensions be given in Gauss
coordinates. Suppose that $e^{-2Ht}g_{ab}$, $e^{2Ht}g^{ab}$,
$e^{2Ht}\sigma^a{}_b$ and their spatial derivatives of all orders are
bounded. Then the solution has an asymptotic expansion of the form given
in Theorem 3. The expansion remains valid when differentiated term
by term to any order.

\noindent
{\bf Proof} The Hamiltonian constraint can be used to express $\tr k$
in terms of the scalar curvature $R$, $\sigma^a{}_b$ and $\Lambda$, giving
\begin{equation}
\tr k=-\left[\frac{n}{n-1}\left(-R+\sigma^a{}_b\sigma^b{}_a\right)+n^2 H^2
\right]^{1/2}
\end{equation}
It follows from the assumptions of the theorem that 
$\tr k=-nH+O(e^{-2Ht})$ and that this relation may be differentiated
term by term with respect to the spatial variables. Now
\begin{equation}
\d_t(e^{-2Ht}g_{ab})=-2e^{-2Ht}g_{ac}(k^c{}_b+H\delta^c_b)
\end{equation}
The right hand side of this expression is $O(e^{-2Ht})$ and so there
is some $g^0_{ab}$ such that 
\begin{equation}
e^{-2Ht}g_{ab}=g^0_{ab}+O(e^{-2Ht})
\end{equation}
and corresponding relations hold for spatial derivatives of all orders.
Using the evolution equations it can be seen that these relations can also 
be differentiated repeatedly with respect to time. The proof now procedes by 
induction. The inductive hypothesis is as follows. There exist coefficients 
$(g_{ab})_{m,l}$ and $(k^a{}_b)_{m,l}$,$0\le m\le M$ such that 
\begin{eqnarray}
g_{ab}&=&e^{2Ht}(\sum_{m=0}^M\sum_{l=0}^{L_m} (g_{ab})_{m-2,l}t^le^{-mHt}
+\bar g_{ab})=[g_{ab}]_M+e^{2Ht}\bar g_{ab} \\
k^a{}_b&=&\sum_{m=0}^M\sum_{l=0}^{L_m} (k^a{}_b)_{m,l} t^le^{-mHt}
+\bar k^a{}_b=[k^a{}_b]_M+\bar k^a{}_b
\end{eqnarray}
where $\bar g_{ab}$ and $\bar k^a{}_b$ are $O(e^{-(M+\epsilon)Ht})$ and 
similar asymptotic expansions hold for all derivatives of these 
quantities. Here $\epsilon$ is a constant belonging to the 
interval $(0,1)$. The inductive hypothesis is satisfied for $M=1$.
If these expressions are substituted into the Einstein
equations then the expansion coefficients written explicitly
satisfy the same relations as in the analysis of formal power series 
solutions carried out above. It is convenient to write the evolution
equations in the following form:
\begin{eqnarray}
\d_t \hat g_{ab}&=&-2\hat g_{ac}\sigma^c{}_b-(2/n)(\tr k+nH)\hat g_{ab}       
\label{support1}     \\
\d_t\sigma^a{}_b+nH\sigma^a{}_b&=&(\tr k+nH)\sigma^a{}_b+\tilde R^a{}_b 
\label{support2}\\
\d_t (\tr k+nH)+2nH(\tr k+nH)&=&(\tr k+nH)^2+R\label{support3}
\end{eqnarray}
where $\hat g_{ab}=e^{-2Ht}g_{ab}$.
Using the inductive hypothesis it follows that if each quantity $Q$ in
these equations is replaced by the corresponding quantity $[Q]_{M+1}$ then
equality holds up to a remainder of order $e^{-(M+1+\epsilon)Ht}$
in (\ref{support2}) and (\ref{support3}). Using this information shows
that the corresponding statement holds in (\ref{support1}) with a
remainder of order $e^{-(M-1+\epsilon)Ht}$.
Thus the quantities $[Q]_{M+1}$ satisfy a system of the type occurring in 
Lemma 1. It follows from that lemma that the inductive hypothesis is 
satisfied with $M$ replaced by $M+1$.

\vskip 10pt\noindent
In \cite{lim} results similar to those of this section were obtained
using a different coordinate system. The time coordinate used there 
satisfies the condition that that the lapse function is proportional
to the inverse of the mean curvature of its level surfaces. This 
means that the foliation of level surfaces is a solution of the inverse 
mean curvature flow, a fact which raises serious doubts whether such 
coordinates exist in forever expanding cosmological spacetimes, as will 
now be explained. The inverse mean curvature flow for hypersurfaces is 
defined by the condition that a hypersurface flows with a speed 
equal to the inverse of its mean curvature in the normal direction.
In the case of a Riemannian manifold it was used in the work of Huisken
and Ilmanen \cite{huisken} on the Penrose inequality. For spacelike 
hypersurfaces in a Lorentzian manifold it was studied in \cite{holder}.
If a spacelike hypersurface with positive expansion (i.e., in the convention
used here, with $\tr k<0$) is given then there is a local solution of the
inverse mean curvature flow in the contracting direction. Moreover, under
reasonable assumptions on the nature of the singularity, there is a global
solution. In the expanding direction, in contrast, the equation is 
backward parabolic and it is to be expected that there is no local solution 
for general initial data, i.e. for a general starting hypersurface. This
is an analogue of the fact that the heat equation cannot be solved 
backwards in time. 

\section{Relations to conformal infinity}

There is a relation between the expansions discussed in the last two 
sections and the concept of conformal infinity. In this section only
the Einstein vacuum equations are considered. Define $T=H^{-1}e^{-Ht}$.
Then spacetime metric corresponding to (\ref{logformal}) becomes
\begin{equation}
(HT)^{-2} [-dT^2+(g^0_{ab}+\sum_{m=0}^\infty\sum_{l=0}^{L_m}(g_{ab})_{m,l}
(-1)^lH^{-l}(\log (HT))^l(HT)^m)]
\end{equation}
It is conformal to a metric which is non-degenerate at $T=0$ and is
written in Gauss coordinates. If there are no non-vanishing coefficients
with $l>0$ the conformal metric (or unphysical metric) is smooth at
$T=0$. This is for instance the case when $n$ is odd.

In the case $n=3$ Friedrich \cite{friedrich86}, \cite{friedrich91} has used 
conformal techniques to prove results which, as shown in the following, 
imply that spacetimes evolving from initial data 
close to standard initial data for de Sitter space indeed have asymptotic 
expansions of the type presented in the last section. The method used,
based on the conformal method, is only known to work in the case $n=3$.
The occurence of logarithms in the expansions for even values of $n$ cast 
doubt on the possibility of implementing an analogous procedure in that case.
There are also problems for $n=3$ if matter is present. For conformally
invariant matter fields the method can be used but for other types of
matter, e.g. a perfect fluid with linear equation of state, there is
no straightforward way of doing this. The non-integer powers occurring
in the formal expansions for this case make the application of the method
problematic. Note, however, that a similar problem has been overcome
in the study of isotropic singularities \cite{anguige99}. 

Consider the de Sitter solution with a slicing by intrinsically flat
hypersurfaces, as described in Section 2, with the slicing being
given by the hypersurfaces of constant $t$. We may assume for 
convenience that the solution has been identified in a way which is 
periodic in the spatial coordinates. Consider initial data which is
a small perturbation of the initial data induced by this model solution
on a hypersurface of constant time. The smallness can be measured in
the sense of uniform convergence of a function and its derivatives of
all orders. Then, according to section 9 of \cite{friedrich91}, the 
perturbed  solution has a Cauchy development which is asymptotically
simple in the future. This means that the solution $g_{\alpha\beta}$
is conformal to a metric 
$\tilde g_{\alpha\beta}=\Omega^2 g_{\alpha\beta}$ with $\Omega>0$
in such a way that $g_{\alpha\beta}$ and $\Omega$ have smooth
extensions through a hypersurface where $\Omega$ vanishes. We may
choose coordinates in the unphysical metric in the following way.
Set $\tilde T=\Omega$ and choose the spatial coordinates to be constant 
along the curves orthogonal to the hypersurfaces of constant $\Omega$.
In these coordinates the conformal metric takes the form
\begin{equation}
-H^{-2}\alpha^2 d\tilde T^2+\tilde g_{ab} dX^a dX^b
\end{equation}
where $\alpha$ is a function of $\tilde T$ and $X^a$. 
The condition $-3\nabla_\alpha\Omega\nabla^\alpha\Omega=\Lambda$ 
(see Lemma 9.2 of \cite{friedrich91}) implies that $\alpha=1$
for $\tilde T=0$. In order to compare this with the expansions in the 
previous sections we need to transform to Gauss coordinates with 
respect to the physical metric $g_{\alpha\beta}$. As a first step
let $\tilde T=e^{-HT}$. Then the physical metric becomes
\begin{equation}\label{rawmetric}
-\alpha^2 dT^2+g_{ab} dX^a dX^b
\end{equation}
with $g_{ab}=e^{2HT}\tilde g_{ab}$. The following lemma shows that Gauss 
coordinates of a suitable kind can be introduced. The hypotheses make use 
of the following inequalities
\begin{eqnarray}
|g_{ab}|&\le& Ce^{2HT}\label{ineq1}                               \\
|\alpha^{-1}|+|\Gamma^a_{bc}|&\le& C\label{ineq2}                  \\
|\alpha-1|+|\d_T\alpha|+|\d_a\alpha|&\le& Ce^{-HT}\label{ineq3}   \\
|\tilde k^a{}_b|+|\tr k +3H|+|g^{ab}|&\le& Ce^{-2HT}\label{ineq4}
\end{eqnarray}
The metric (\ref{rawmetric}) above satisfies inequalities of this type 
together with corresponding inequalities for spatial derivatives 
of all orders. In fact the estimates for $\tilde k^a{}_b$ and $\tr k+H$
are only obviously satisfied with the bound $Ce^{-HT}$. However this can
be improved by using equations (\ref{support2}) and (\ref{support3})
at the end of the last section, or rather their equivalents in the 
presence of a non-trivial lapse function.

\vskip 10pt\noindent
{\bf Lemma 2} Consider a metric of the form (\ref{rawmetric}) 
on a time interval $[T_0,\infty)$ and assume that there is a constant $C>0$ 
such that the inequalities (\ref{ineq1})-(\ref{ineq4}) are satisfied, 
together with the corresponding inequalities for spatial
derivatives of all orders. Then for $T_0$ sufficiently large there exists 
a Gaussian coordinate system based on the hypersurface $T=T_0$ which is 
global in the future. The transformed metric satisfies the hypotheses of 
Theorem 5.

\noindent
{\bf Proof} To construct Gaussian coordinates it is necessary to 
analyse the equations of timelike geodesics. In 3+1 form these are
\begin{eqnarray}
\frac{d^2 T}{d\tau^2}&+&\alpha^{-1}\d_T \alpha\left(\frac{dT}{d\tau}\right)^2
+2\alpha^{-1}\nabla_a\alpha\frac{dX^a}{d\tau}\frac{dT}{d\tau}+\alpha^{-1}
k_{ab}\frac{dX^a}{d\tau}\frac{dX^b}{d\tau}=0    \\
\frac{d^2 X^a}{d\tau^2}&+&\alpha\nabla^a\alpha\left(\frac{dT}{d\tau}\right)^2
-\frac{2}{3}\alpha (\tr k)\frac{dX^a}{d\tau}\frac{dT}{d\tau} \nonumber\\
&-&2\alpha\tilde k^a{}_b\frac{dX^b}{d\tau}\frac{dT}{d\tau}
+\Gamma^a_{bc}\frac{dX^b}{d\tau}\frac{dX^c}{d\tau}=0
\end{eqnarray}
It is helpful for the following analysis to rewrite one of the terms:
\begin{eqnarray}
-\frac{2}{3}\alpha (\tr k)\frac{dX^a}{d\tau}\frac{dT}{d\tau}
&=&2H\frac{dX^a}{d\tau}+2H\frac{dX^a}{d\tau}\left(\frac{dT}{d\tau}-1\right)
\nonumber\\
&-&\frac{2}{3}(\tr k+3H)\frac{dX^a}{d\tau}\frac{dT}{d\tau}
-\frac{2}{3}(\alpha -1) (\tr k)\frac{dX^a}{d\tau}\frac{dT}{d\tau}
\end{eqnarray}
These equations are to be solved for functions $T(\tau,x^b)$ and 
$X^a(\tau,x^b)$ with initial values $T=T_0$, $dT/d\tau=1$, $X^a=x^a$ and 
$dX^a/d\tau=0$ at $\tau=T_0$. Strictly speaking Gaussian coordinates based 
on $T=T_0$ would differ from this by a time translation by $T_0$ but it is 
convenient here to work with this slight modification. Consider now a 
solution of these equations on an interval $[T_0,\tau^*)$ and suppose for 
later convenience that $T_0\ge 0$. There is a $\tau^*>T_0$
for which a solution does exist. We assume that on this interval 
$|dX^a/d\tau|\le Ce^{-2H\tau}$ and that $|dT/d\tau-1|<\epsilon$ for some
$\epsilon\in (0,1/3)$. For given $C$ and $\epsilon$ there exists an interval of
this kind. On this interval $e^{-T}\le e^{-\epsilon\tau_0}
e^{-(1-\epsilon)\tau}$, $e^{T-\tau}\le e^{\epsilon (\tau-\tau_0)}$ and 
inequalities of the following form hold, where $C'$ is a positive constant 
depending only on $C$ and $\epsilon$.
\begin{eqnarray}
\frac{d^2 T}{d\tau^2}&=& f(\tau),\ \ |f(\tau)|\le C'e^{-(1-\epsilon)H\tau}  \\
\frac{d^2 X^a}{d\tau^2}+2H\frac{dX^a}{d\tau}
&=&g(\tau), \ \ |g(\tau)|\le C' e^{-2H\tau}
\end{eqnarray}
It follows from the first of these that
\begin{equation}
|dT/d\tau-1|\le C'e^{-(1-\epsilon)HT_0}
\end{equation}
For $T_0$ large enough this strictly improves on the estimate originally
assumed for $dT/d\tau-1$. For small $\epsilon$ the quantities 
$dX^a/d\tau$ can be seen to decay exponentially with an exponent which
is as close as desired to $-2$. The fact that we are dealing with timelike
geodesics parametrized by proper time leads to the relation
\begin{equation}
-1=-\alpha^2\left(\frac{dT}{d\tau}\right)^2
+g_{ab}\frac{dX^a}{d\tau}\frac{dX^b}{d\tau}
\end{equation}
This implies that $|dT/d\tau-1|=O(e^{-HT})$. Putting this back into the 
evolution equation for $dX^a/d\tau$ shows that it is $O(e^{-2H\tau})$.
By choosing $T_0$ large enough the decay estimate for this quantity
is recovered and in fact strengthened. Consideration of the longest
time interval on which the original inequality holds shows that
$\tau^*=\infty$. The estimates we have derived up to now hold globally.
The estimate for $dT/d\tau$ obtained above implies that there are
positive constants $C_1$ and $C_2$ such that $C_1\tau\le T\le C_2\tau$.
Hence in estimates we can replace $e^{-T}$ by $e^{-\tau}$ if desired.

Next we would like to obtain corresponding estimates for the spatial
derivatives of $dT/d\tau$ and $dX^a/d\tau$ of all orders. Consider the
result of differentiating the geodesic equations with respect to the
spatial variables. This leads to estimates of the form
\begin{eqnarray}
\frac{d}{d\tau}\left(\frac{\partial^2 T}{\partial\tau\partial x^a}\right)
&=& f_a(\tau),\ \ |f_a(\tau)|\le C'e^{-H\tau}  \\
\frac{d}{d\tau}\left(\frac{\partial^2 X^a}{\partial\tau\partial x^c}\right)
+2H\frac{\partial^2 X^a}{\partial\tau\partial x^c}
&=&g^a_c(\tau), \ \ |g^a_c(\tau)|\le C' e^{-2H\tau}
\end{eqnarray}
This allows us to show that the first order spatial derivatives of
the key quantities satisfy the estimates analogous to those satisfied by 
the quantities themselves. The same argument can be applied to estimate 
spatial derivatives of any order inductively. Now all the desired 
information about existence and decay of $T$ and $X^a$ has been obtained.
It remains to show that they form a coordinate system. This follows
from the fact that the initial values of $\partial T/\partial\tau$,
$\partial X^a/\partial\tau$ and $\partial X^a/\partial x^b$ are one,
zero and $\delta^a_b$ respectively and the exponential decay of their
time derivatives which has already been proved. This completes the proof 
of the lemma. 

\vskip 10pt
Combining Lemma 2 and Theorem 5 shows that the spacetimes constructed by
Friedrich admit global Gaussian coordinates in which they have an
asymptotic expansion of the form of (\ref{formal}). Hence any initial
data close to that for de Sitter on a flat hypersurface evolves into
a solution having an asymptotic expansion of the form given by Starobinsky.

\section{The spatially homogeneous case}

This section is concerned with spatially homogeneous solutions of the
vacuum Einstein equations in $n+1$ dimensions with positive cosmological
constant. It is assumed that the spatial homogeneity is defined by a Lie 
group $G$, supposed simply connected, which acts simply transitively. We 
restrict to spacetimes such that all left invariant Riemannian metrics on 
$G$ have non-positive scalar curvature. In the case $n=3$ this corresponds 
to Bianchi types I to VIII.  Information on the case $n=4$ can be found
in \cite{hervik}. It will be shown that the spacetimes of the type just
specified have asymptotic expansions with all the properties of the 
formal expansions in Theorem 3.

A spatially homogeneous spacetime of the type being considered can be written 
in the form
\begin{equation} 
-dt^2+g_{ij}(t)e^i\otimes e^j
\end{equation}
where $\left\{e^i\right\}$ is a left invariant frame on the Lie group 
$G$. Basic information about the
asymptotics of these spacetimes in 3+1 dimensions are given by Wald's 
theorem \cite{wald}, which provides information on the behaviour of the 
second fundamental form as $t\to\infty$. This can easily be generalized
to the present situation. Using the condition on the sign of $R$, the
Hamiltonian constraint implies that on any interval where a solution
exists $(\tr k)^2\ge\frac{2n\Lambda}{n-1}=(nH)^2$. Combining the Hamiltonian 
constraint with the evolution equation for $\tr k$ gives
\begin{equation}\label{diffineq}
\d_t (\tr k)\ge \frac1n (\tr k)^2-\frac2{n-1}\Lambda
\end{equation}
In particular $\tr k$ is non-decreasing. These facts together show
that $\tr k$ is bounded. Now it will be shown that $\tr k\to -nH$
as $t\to\infty$. For
\begin{eqnarray}
\d_t(\tr k+nH)&\ge&\frac1n (-\tr k+nH)(-\tr k-nH)    \\
&\ge& -2H(\tr k+nH)
\end{eqnarray}
It follows that $\tr k=-nH+O(e^{-2Ht})$. Using the Hamiltonian constraint
then gives $\sigma^i{}_j\sigma^j{}_i=O(e^{-2Ht})$. This bound can be
used to get information on $\sigma_{ij}$ as in \cite{rendall94}. Then 
it is possible to proceed exactly as in the proof of Proposition 2 in 
\cite{lee} to show that $e^{-2Ht}g_{ij}$, $e^{2Ht}g^{ij}$ and 
$e^{Ht}\sigma^i{}_j$ are bounded. Then equation (58) can be used as
in the previous section to improve the last statement to the boundedness
of $e^{2Ht}\sigma^i{}_j$. The fact that $g_{ij}$, $g^{ij}$ and $k_{ij}$ are
bounded on any finite time interval implies that the solution exists
globally in time.

We are now in a situation very similar to that of Theorem 5. However the 
estimates we have are expressed in term of frame components. Choosing a 
coordinate system on some subset of the Lie group $G$ with compact closure 
will give us uniform asymptotic expansions for the components in that 
coordinate system. Conversely uniform asymptotic expansions for the 
components in a coordinate system of this type give corresponding asymptotic 
expansions for the frame components. In this case we will say that the 
asymptotic expansions are locally uniform. If an expansion of this type holds 
for a given quantity it also holds for all spatial derivatives in the 
coordinate representation.

The proof of Theorem 5 uses only arguments which are pointwise in space and
so it generalizes immediately to the case of locally uniform asymptotic
expansions. It can be concluded that for a spatially homogeneous spacetime
of the type under consideration locally uniform asymptotic expansions
of generalized Starobinsky type are obtained. Restricting to a coordinate
domain with compact closure uniform asymptotic expansions are obtained.
In general these expansions will contain logarithmic terms. Consider for
instance the case of the Lie group $H\times \bf R$ where $H$ is a 
three-dimensional Lie group of Bianchi type other than IX. Let the
spacetime be such that the spatial metric at each time is the product of 
a metric on $H$ with one on $\bf R$. This is consistent with the constraint
equations. For instance the initial data can be chosen to be the product
of data on $H$ with trivial data on $\bf R$. The data are invariant under
reflection in $\bf R$ and this property is inherited by the solutions. 
Suppose that $H$ admits 
no metric of vanishing scalar curvature. This is the case for every Bianchi 
type except I and VII${}_0$. Then the metric $g^0_{ab}$ corresponding to this 
solution has non-vanishing scalar curvature and is not an Einstein metric. 
Hence logarithmic terms are unavoidable. 

\section{Fuchsian analysis}

Fuchsian systems are a class of singular equations which can be used to
prove the existence of solutions of certain partial differential equations
with given asymptotics \cite{kichenassamy}, \cite{kr}, \cite{rendall03}. It 
will be shown that Fuchsian methods allow the construction of solutions of the 
vacuum Einstein equations with positive cosmological constant in any number 
of dimensions which have asymptotic expansions of the type given in Section 2
and depend on the same number of free functions as the general solution.

Before coming to the specific problem of interest here some general
facts about Fuchsian equations will be recalled. The form of the 
equations is
\begin{equation}\label{fuchs}
t\d u/\d t+N(x)u=tf(t,x,u,u_x)
\end{equation}
Here $x$ denotes the spatial coordinates collectively and $u_x$ the 
spatial derivatives of the unknown $u(t,x)$. The matrix $N$ and the
function $f$ are required to satisfy certain regularity conditions
and $N$ is required to satisfy a positivity condition. There are forms
of the regularity condition adapted to smooth and to analytic functions.
The version adapted to analytic fucntions will be used in the following
since it is the one where the most powerful theorems are available.
For the precise definition of regularity see \cite{ar}, where a
corresponding definition of regularity of solutions is also given.
Roughly speaking, regularity means that the functions concerned are
continuous in $t$ and analytic in $x$ and vanish in a suitable way as 
$t\to 0$. Consider now an ansatz of the form
\begin{equation}
u(t,x)=\sum_{m=0}^\infty\sum_{l=0}^{L_m} u_{m,l}t^m(\log t)^l
\end{equation}
By analogy with what was done in Section 2 we can ask whether the equation 
(\ref{fuchs}) has a formal series solution of this kind. Suppose that this
is the case. Fix $M\ge 0$. Then there exist coefficients $u_{m,l}$
such that $\bar u=u-\sum_0^M\sum_0^{L_m} u_{m,l}t^m(\log t)^l$ satisfies
\begin{equation}
t\d \bar u/\d t+N(x)\bar u=t^{M+\epsilon}f_M(\bar u)
\end{equation}
for a regular function $f_M$ and a constant $\epsilon>0$ together with the 
corresponding relations obtained by differentiating term by term with respect 
to the spatial coordinates
any number of times. Let $v=t^{-M}\bar u$. Then
\begin{equation}
t\d v/\d t+(N(x)+MI)v=t^\epsilon g(t,x,v,v_x)
\end{equation}
for a regular function $g$. Introducing $t^{\epsilon}$ as a new time
variable, we obtain an equation of the form (\ref{fuchs}).
If we assume that $N(x)$ is bounded then by
choosing $M$ large enough it can be ensured that the matrix $N(x)+MI$
is positive definite. Assuming that $f$ and $N$ are regular in the 
analytic sense the existence theorem of \cite{ar} implies
the existence of a unique regular solution $v$ vanishing at $t=0$.
Expressing $u$ in terms of $v$ gives a solution of the original
equation which has the given asymptotic expansion up to order $M$.

Consider now the slightly more general equation
\begin{equation}\label{genfuchs}
t\d u/\d t+N(x)u=tf(t,x,u,u_x)+h(u)
\end{equation}
In order to have a consistent formal power series solution suppose that
for some functions $u_{0,0}$ and $u_{1,0}$ we have $Nu_{0,0}=h(u_{0,0})$ and 
\begin{equation}
(N+I-Dh(u_{0,0}))u_{1,0}=f(0,u_{0,0})
\end{equation} 
Here $Dh$ denotes the derivative of 
$h$ as a map between Euclidean spaces. Suppose further that the equation 
admits a formal power series solution with coefficients $u_{0,0}$ and 
$u_{1,0}$ and $L_0=L_1=0$. If these conditions hold then $u$ satisfies
the original equation if $v=u-u_{0,0}-u_{1,0}t$ satisfies a Fuchsian 
system and vanishes at the origin. Thus an existence theorem is obtained.

To make contact with the Einstein equations we start with the equations 
(\ref{support1})-(\ref{support3}) and set $\tau=e^{-Ht}$. Then an
equation of the form (\ref{genfuchs}) is obtained, with
$u=(\hat g_{ab},\sigma^a{}_b,\tr k+nH)$. If it is assumed that
the variables $\sigma^a{}_b$ and $(\tr k+nH)$ vanish at $\tau=0$ then
the consistency conditions on $u_{0,0}$ and $u_{1,0}$ are satisfied. The 
fact that 
consistent formal expansions were shown to exist in Section 2 allows the
above procedure to be carried through. If the data $A_{ab}$ and $B_{ab}$
are chosen to be analytic then this gives an existence theorem
for the Einstein evolution equations with $A_{ab}$ and $B_{ab}$ prescribed 
as in Theorem 3. In this context it is important to note that if
$A_{ab}$ and $B_{ab}$ are analytic all the coefficients in the formal
expansions whose existence is asserted in Theorem 3 are also analytic.
In order to see that a solution of the Einstein
equations is obtained it suffices to show that the constraint equations
are satisfied. Note that it follows from the results of Section 2 that
the constraint quantities vanish to all orders at $\tau=0$ but since 
the solution is not analytic at $\tau=0$ this does not suffice to conclude
that the constraint quantities vanish everywhere. To see that they do we 
need to write the consistency conditions (\ref{hamconsistency}) and 
(\ref{momconsistency}) in Fuchsian form. Using the fact that the Einstein 
evolution equations are satisfied, and introducing $\tilde C_a=e^{-tH}C_a$,
these equations can be written as
\begin{eqnarray}
\d_t C+2HC&=&2(\tr k+H)C-2e^{Ht}\nabla^a C_a        \\
\d_t \tilde C_a+2H\tilde C_a&=&(\tr k+H)\tilde C_a-(1/2)e^{-Ht}
\nabla_a\tilde C
\end{eqnarray}
Setting $\tau=e^{-Ht}$ gives a system of the form (\ref{fuchs}) and 
since $C$ and $\tilde C_a$ tend to zero as 
$\tau\to 0$ both of these quanitities vanish as a consequence of the 
uniqueness theorem for Fuchsian systems and the constraints are 
satisfied. The solution of the Einstein equations has an asymptotic expansion 
of the form given
in Theorem 3 truncated at any given finite order. Applying Theorem 5 shows 
that this solution has an asymptotic expansion of this form to all orders.
The results obtained can be summed up as follows:

\vskip 10pt\noindent
{\bf Theorem 6} Let $A_{ab}$ be an analytic $n$-dimensional Riemannian metric
and $B_{ab}$ an analytic symmetric tensor which satisfies 
$A^{ab}B_{ab}=Z(A)$ and $\nabla^a B_{ab}=\tilde Z_b(A)$, where the
covariant derivative is that associated to $A$. Then there exists an
analytic solution of the vacuum Einstein equations with
an asymptotic expansion of the form (\ref{logformal}) with $g^0_{ab}=A_{ab}$ 
and $(g_{ab})_{n-2,0}=B_{ab}$. The expansion may be differentiated term by
term with respect to the spatial variables as often as desired.

\section{The wave equation on de Sitter spacetime}

In Section 4 it was shown that initial data for the vacuum Einstein equations
in 3+1 dimensions close to that for de Sitter space evolve to give a 
spacetime with asymptotics of Starobinsky type. It has not yet proved 
possible to obtain the analogous statement in higher dimensions. What is
missing are suitable energy estimates. In this section it will be shown how 
a simpler model problem can be treated. This is the case of the wave equation
$\nabla_\alpha\nabla^\alpha\phi=0$ on (the higher dimensional analogue of) de
Sitter space. The spacetime metric in this case is
\begin{equation}
ds^2=-dt^2+e^{2Ht}((dx^1)^2+\dots+(dx^n)^2)
\end{equation}
Written out explicitly in coordinates the wave equation takes the form:
\begin{equation}
\d_t^2\phi+nH\d_t\phi=e^{-2Ht}\Delta\phi
\end{equation}
where $\Delta$ is the Laplacian of the flat metric.

Consider the ansatz for formal solutions of the equations
\begin{equation}
\sum_{m=0}^\infty (A_m(x)e^{-mHt}+B_m(x)te^{-mHt})
\end{equation}
Substituting this into the equation and comparing coefficients gives
\begin{eqnarray}
m(m-n)H^2A_m-(2m-n)HB_m&=&\Delta A_{m-2} \\
m(m-n)H^2B_m&=&\Delta B_{m-2}
\end{eqnarray}
For any $n$ it is true that $B_0=A_1=B_1=0$. In the case that $n$ is odd
assume that the coefficients $B_m$ vanish. Then 
$\Delta A_{m-2}=H^2m(m-n)A_m$ for all $m\ge 2$. Then it follows from $A_1=0$
that $A_{2k+1}=0$ for all integers $k$ with $2k+1<n$. The coefficients 
$A_{2k+1}$ with $2k+1>n$ are determined by $A_n$. The coefficients 
$A_{2k}$ are determined by $A_0$. There are no further relations to be
satisfied and so $A_0$ and $A_n$ parametrize the general solution.
If the coefficients $B_m$ are not assumed to be zero it can be shown 
that they must vanish for $n$ odd. For $n$ even we have 
$A_{2k+1}=B_{2k+1}=0$ for every positive integer $k$. If $2k<n$ then
$B_{2k}=0$. For $2k<n$ the coefficients $A_{2k}$ are determined successively
by $A_0$. Also $B_n$ is determined by $A_0$. Then $A_{2k}$ and $B_{2k}$
are determined for all $k$ with $2k>n$ in terms of the coefficients already
determined. Thus the general solution can be parametrized by $A_0$ and $A_n$,
just as in the case $n$ odd. The difference is that the series obtained
contains terms which are multiples of $te^{-mHt}$.

Let $E=\int (\d_t\phi)^2+e^{-2Ht}|\nabla\phi|^2$. Differentiating
with respect to $t$ and integrating by parts gives the relation
$dE/dt=-2HE$. We can differentiate the equation through with
respect to a spatial coordinate and repeat the argument. This shows
that all Sobolev norms of $e^{Ht}\d_t\phi$ and $\nabla\phi$ are
bounded. By the Sobolev embedding theorem they and all their spatial
derivatives satisfy corresponding pointwise bounds. Thus the spatial 
derivatives of $\phi$ are bounded while its time derivative decays like 
$e^{-Ht}$. As a consequence $\phi(t,x)=\phi_0(x)+O(e^{-Ht})$ for some 
function $\phi_0$. Comparing with the formal solutions already obtained we
see that these estimates are not likely to be sharp. The equation for
$\phi$ is equivalent to the system
\begin{eqnarray}
\d_t\phi&=&\psi                \\
\d_t\psi+nH\psi&=&e^{-2Ht}\Delta\phi
\end{eqnarray}
Starting with the basic information on the asymptotic behaviour of $\phi$
we already have the method of proof of Theorem 5 can be applied to this
system. The result is that any solution has an asymptotic expansion of
the type derived on a formal level above.

\section{Acknowledgements} I thank Hans Ringstr\"om for discussions on
the subject of this paper and Gary Gibbons for pointing me to relevant work 
on positive definite metrics.

\end{document}